\documentclass[a4paper]{spie}  

\usepackage{amsmath,amsfonts,amssymb}
\usepackage{graphicx}
\usepackage[colorlinks=true, allcolors=blue]{hyperref}

\title{Monolithic interferometric modules for multi-axis coordinate positioning with sub-nanometre precision}

\newenvironment{myitemize}
{ 
\vspace{-2.5mm}
\begin{itemize}
    \setlength{\itemsep}{0pt}
    \setlength{\parskip}{0pt}
    \setlength{\parsep}{0pt}     }
{ \end{itemize} 
    \vskip -2mm} 

\author{Simon Rerucha}
\author{Ondrej Cip}
\author{Miroslava Hola}
\author{Josef Lazar}
\author{Bretislav Mikel}

\affil{\textit{ISI Brno, CZ} -- Institute of Scientific Instruments of the Czech Academy of Sciences,\\ Kralovopolska 147, 612 00 Brno, Czech Republic}

\authorinfo{Contact: res@isibrno.cz, mikel@isibrno.cz}

\begin{document} 
\maketitle

\begin{abstract}

We report on developing, characterizing, and verifying a compact, monolithic laser interferometric assembly designed for high-precision two- and three-axis displacement measurements in coordinate positioning systems. 
The design targets OEM integration into advanced precision motion platforms, including nanometrology instruments, semiconductor manufacturing equipment, and ultra-high vacuum (UHV) environments. 
Using a single laser source, the assembly integrates multiple interferometers into a monolithic L-shaped base frame, enabling sub-nanometer periodic error in X-Y motion systems while minimizing geometric and thermal instabilities. 
The pre-aligned and pre-adjusted architecture simplifies integration, enhances long-term stability, and ensures consistent metrological performance. 
The system's verification protocol employs quadrature phase analysis and systematic error metrics to characterize performance and optimise assembly. 
The results indicate sub-nanometer measurement capability and suggest the system's suitability for scalable implementation in advanced coordinate metrology applications.

\end{abstract}


\textit{Note: This is a preprint rendition of a conference paper S. Rerucha et al, Monolithic interferometric modules for multi-axis coordinate positioning with sub-nanometre precision, Proc. SPIE, vol. 13698, pp. 13698-30, 2025.
}
\section{INTRODUCTION}

The demand for ultra-precise displacement and coordinate measurements continues to grow across a wide range of scientific and industrial domains, including nanometrology, semiconductor manufacturing, and high-resolution positioning systems. Laser interferometry has long served as a fundamental technique for dimensional metrology, offering traceability, sub-nanometer resolution, and high linearity based on well-established interference principles \cite{yang2018review, michelson1887ontherelative}. 

While interferometry provides unmatched resolution, practical implementation for multi-axis coordinate measurement introduces significant engineering challenges \cite{coveney2020review}. Conventional systems often rely on modular interferometric systems, where multiple independent interferometers must be carefully aligned and integrated.
Such an approach suffers from cumulative geometric misalignments, increased sensitivity to thermal gradients, and complex assembly procedures. 
Such limitations become particularly pronounced in OEM scenarios, where a compact form factor, long-term measurement and adjustment stability, vacuum compatibility, and straightforward integration are critical requirements.

To address these limitations, we have developed a compact monolithic interferometric assembly -— referred to as \textit{the vingel} —- which integrates multiple interferometric axes into a single, pre-aligned mechanical and optical structure. The monolithic design ensures stable geometry, reduces thermal drift, simplifies assembly, and enables reproducible performance over time. Using a single laser source, typically a frequency-stabilized He-Ne laser, the system allows simultaneous measurement of two-axis translation and, optionally, the parasitic rotation along the Z-axis with sub-nanometer precision. Furthermore, the design is compatible with ultra-high vacuum (UHV) environments, making it particularly suitable for motion platforms in semiconductor systems.

In the following sections, we present the optical and mechanical configuration of the interferometric assembly (Section \ref{parts}), the implementation of the homodyne receiver architecture (Section \ref{rx}), and the assembly methodology (Section \ref{frame}). Subsequently, we describe the verification and testing procedures (Section \ref{testing}), including quantitative evaluation metrics and sample measurement results (Sections \ref{metrics} and \ref{results}), which collectively demonstrate the system’s ability to achieve high stability, accuracy, and reproducibility suitable for OEM integration into precision positioning systems.

\section{\textit{THE VINGEL}: INTERFEROMETRIC ASSEMBLY FOR COORDINATE MEASUREMENT }
\label{parts}

The three principal parts of the interferometric systems are (i) the interferometers,  (ii) the interferometric receivers and  (iii) the optomechanics of the base frame. Details are presented in the following sections.

\subsection{Interferometers}

Various interferometric configurations are available for displacement measurement along multiple axes. While differential interferometers offer superior accuracy due to common-mode noise rejection \cite{yacoot2000use,pisani2012comparison,rerucha2021compact,rerucha2024interferometer}, their usable range is limited to a few millimetres, and their implementation becomes increasingly complex when extended to multi-degree-of-freedom measurements. Likewise, retroreflector-based designs employing corner cubes exhibit similar range constraints and increased sensitivity to parasitic motions.
Therefore, for the present system, we employ conventional single-ended plane mirror interferometers \cite{steinmetz1990sub}, which offer a favourable trade-off between accuracy, range, and implementation complexity for multi-axis displacement measurements. 

\vskip 1cm 
  \begin{figure} [ht]
   \begin{center}
   \begin{tabular}{c} 
   \includegraphics[width=14cm]{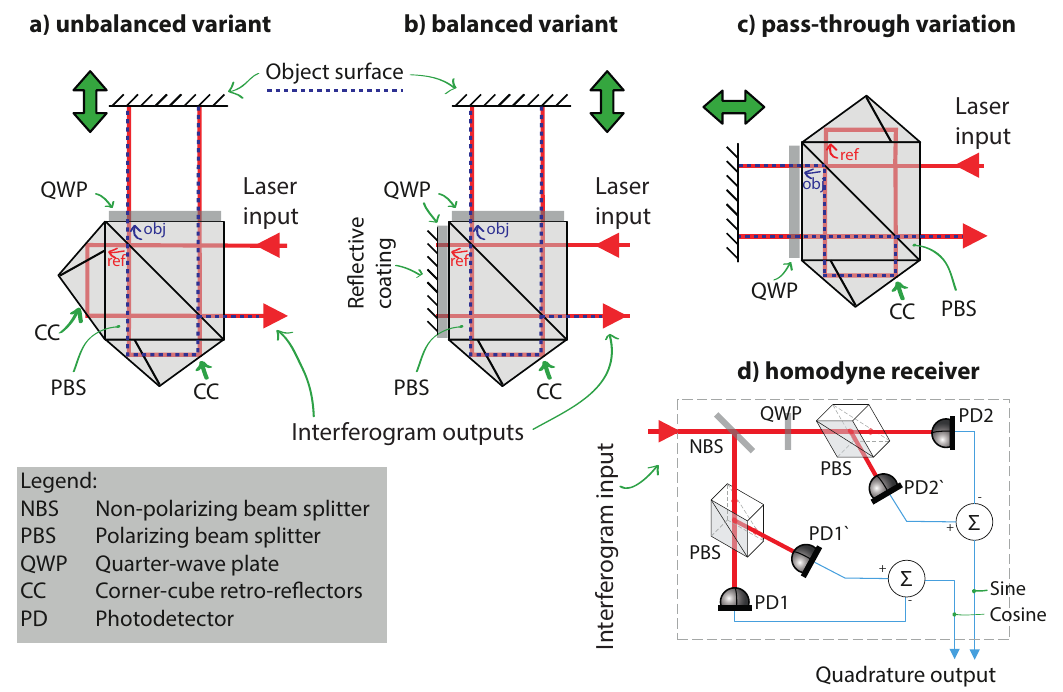}
   \end{tabular}
   \end{center}
   \caption[schema] 
   { \label{fig:schema} The different variants of optical arrangement (adapted from \cite{rerucha2024jbox}): the double-pass interferometer for coordinate measurement in the right-angled variant (a), a modified variant with balanced optical paths (b) and pass-through variant; the conversion of the optical interferogram into electrical signals carries out the four-detector homodyne receiver (d)
}
\end{figure} 

Figure \ref{fig:schema} shows the optical configurations employed. Depending on the application-specific requirements, several variants of the interferometer could be used, for example:

\begin{myitemize}
    \item Double-pass unbalanced arrangement (Figure \ref{fig:schema}a): This configuration offers a minimal optical footprint, making it advantageous for dense multi-axis integration.

    \item Balanced double-pass arrangement (Figure \ref{fig:schema}b): The equalization of optical path lengths minimizes sensitivity to thermal gradients, reducing long-term zero drift [9]. However, this configuration requires tighter manufacturing tolerances to suppress unwanted parasitic reflections and cavity effects.

    \item Pass-through arrangement (Figure \ref{fig:schema}c): This geometry enables multiple interferometric axes with parallel beams, which is essential for three-axis configurations.
\end{myitemize}
    
All the optical components are aligned and bonded either into a compact optical block using optical adhesive or mounted into a dedicated metal housing, ensuring precise relative positioning of optical elements. Once assembled, the interferometric modules maintain fixed alignment, minimizing the need for end-user adjustments.

\subsection{Homodyne Receiver}
\label{rx}

The vingel employs a homodyne quadrature receiver based on a four-photodiode configuration (Figures   \ref{fig:schema}d and \ref{fig:photo}b), providing robust phase detection with high linearity and bandwidth. The interferometric signal emerging from the output contains two orthogonally polarized components, which are separated and processed as follows:

\begin{myitemize}
    \item  The optical signal is split into two channels using a non-polarizing beam splitter.
    \item One channel incorporates a phase retarder adjusted to introduce a $\pi/2$ phase shift between polarization components.
    \item Both channels are directed into polarizing beam splitters oriented at $45\deg$, producing two interferograms with opposite phases.
    \item Each photodiode pair generates differential signals, yielding two quadrature outputs of the form: 
    \begin{equation}
        I_x = K_xcos(\varphi), I_y = K_Ysin(\varphi). 
    \end{equation}
        
\end{myitemize}

The resulting sinusoidal quadrature signals $I_x, I_y$ with amplitudes $K_x, K_y$ are directly usable for downstream displacement computation.

\begin{figure} [ht]
   \begin{center}
   \begin{tabular}{c c} 
   a) \includegraphics[height=3.5cm]{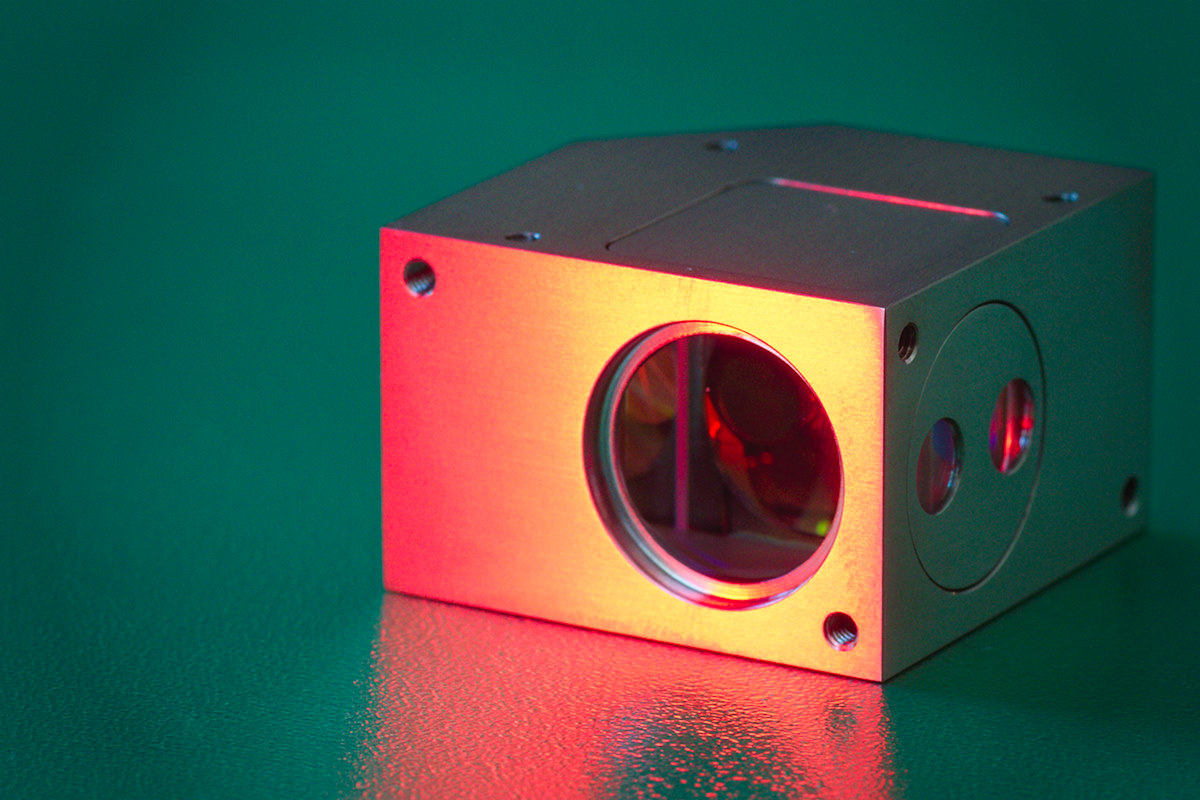} & b) \includegraphics[height=3.5cm]{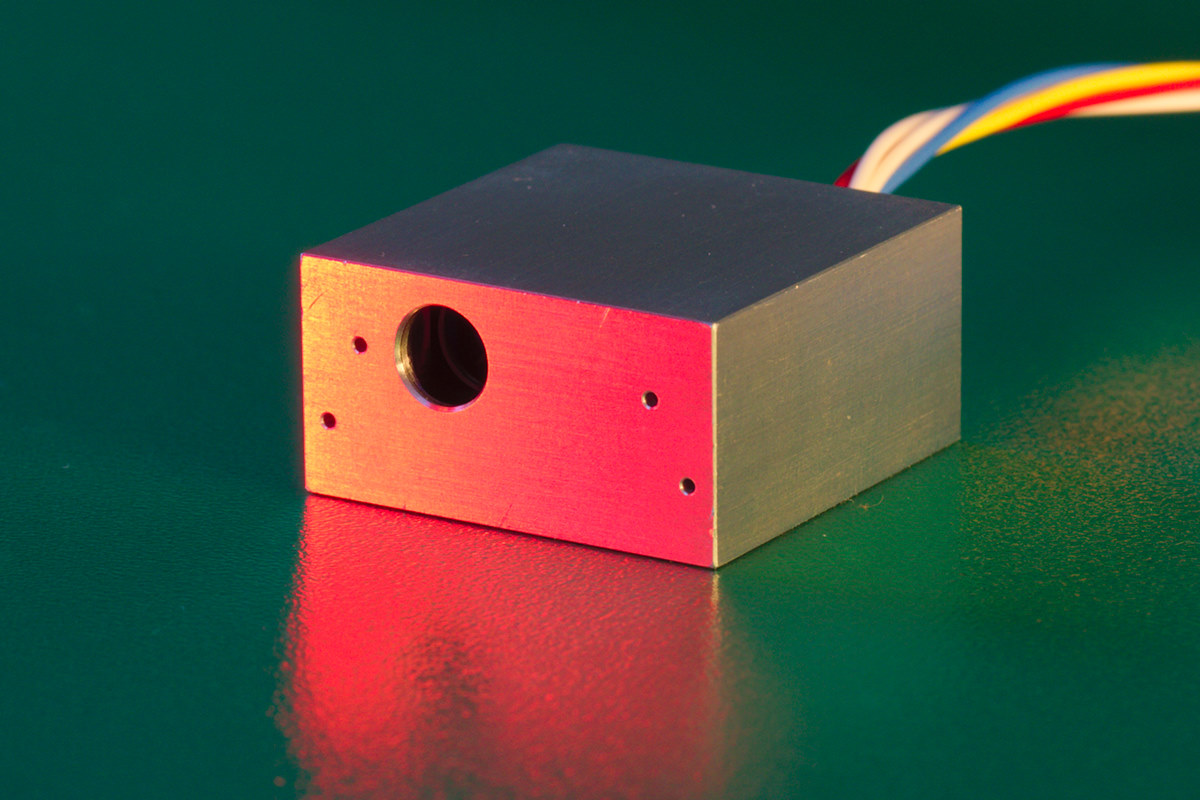}
   \end{tabular}
   \end{center}
   \caption[photo] 
   { \label{fig:photo} 
Assembled components including housing: the double-pass planar interferometer from Figure \ref{fig:schema}a for coordinate measurement (a): the clear aperture left on the left serves for beam input and interferogram outputs, the measurement arm's beams use smaller apertures on the right; the four-detector homodyne receiver unit shown in Figure \ref{fig:schema}d (b)}
\end{figure} 

The homodyne receiver interface provides analogue quadrature outputs, a symmetric power supply interface, and a common ground. While this analogue signal interface may appear unconventional relative to modern digital sensor technologies, it offers minimal latency, wide bandwidth, and compatibility with real-time control systems in high-speed positioning platforms.

\subsection{Baseframe and Assembly}
\label{frame}

\begin{figure} [ht]
   \begin{center}
   \begin{tabular}{c} 
   \includegraphics[height=5.5cm]{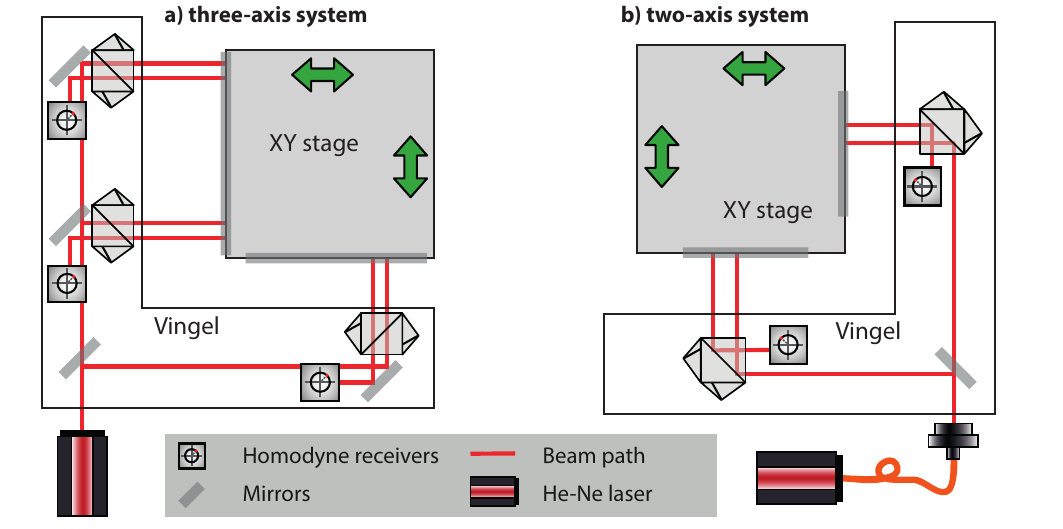} 
   \end{tabular}
   \end{center}
   \caption[photo] 
   { \label{fig:block} 
Block diagram of the interferometric systems: three-axis vingel (a) and two-axis vingel (b)}
\end{figure} 

The L-shaped monolithic baseframe integrates the interferometers, homodyne receivers, and auxiliary beam-steering optics (Figure \ref{fig:block}). Partially reflective mirrors guide and split the incoming laser beam to the individual interferometers and route the returning signals to their respective receivers. All interferometers are aligned for geometric orthogonality (squareness, leveling) and optimized for signal quality during assembly.

The structural components are fabricated from stainless steel or titanium, offering excellent dimensional stability and compatibility with ultra-high vacuum environments due to their low outgassing properties. All optomechanical surfaces are precision ground to achieve high flatness and mechanical integrity.

The laser beam, typically from a single-frequency stabilized He-Ne laser, is delivered via a free-space coupling or through an optical fibre collimator with a polarizing filter added (e.g. Glan-Thompson prism). A typical operating optical power of approximately $\approx 0.5\,\mu$W  for the entire system is sufficient for stable operation. Optical isolators are recommended to suppress unwanted back-reflections that may otherwise degrade system stability.

\section{VERIFICATION TESTING}
\label{testing}

The successful deployment of monolithic interferometric assemblies requires not only precise fabrication and alignment but also systematic verification to ensure consistent performance across multiple units. Unlike one-off laboratory setups, repeatable production demands standardized assembly and testing protocols that can quantify key performance metrics and identify assembly deviations.

The fundamental premise we start from is that the performance of the presented monolithic interferometric systems is strongly related to the assembly and alignment precision of individual optical elements across the system. 

Any imperfections (that could also stem from the quality of the optical elements) would influence the system's optical outputs. So, with proper knowledge of optical phenomena, these imperfections can be detected and, to a certain extent, quantified. 
This information, in turn, provides important feedback through the individual steps along the entire optical assembly, providing a framework for in-production testing. 
At the end of the manufacturing process, the same methodology provides a foundation for verification testing and quality assurance.

The verification process developed for the vingel assembly evaluates both static and dynamic performance, with specific attention to periodic nonlinearity, phase errors, and system stability. The following subsections describe the verification approach, testbed configurations, evaluation metrics, and sample results.

\subsection{Experimental Testbeds and Procedures}

Our experience with assembling the vingels revealed the need for several testbeds for different assembly steps, which include at least: a testbed for individual receivers, a static testbed for the vingel alignment and an actuated testbed for dynamic testing and verification\footnote{The list omits the fixtures needed for glueing and aligning the beam-splitters and other optics.}

\begin{figure} [ht]
   \begin{center}
   \begin{tabular}{c c} 
   a) \includegraphics[height=5cm]{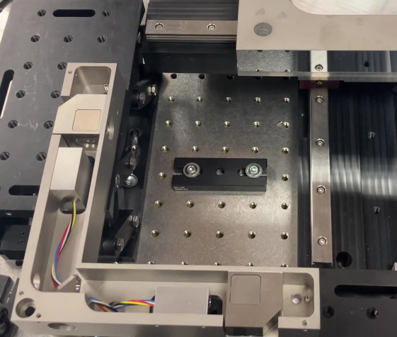} & b) \includegraphics[height=5cm]{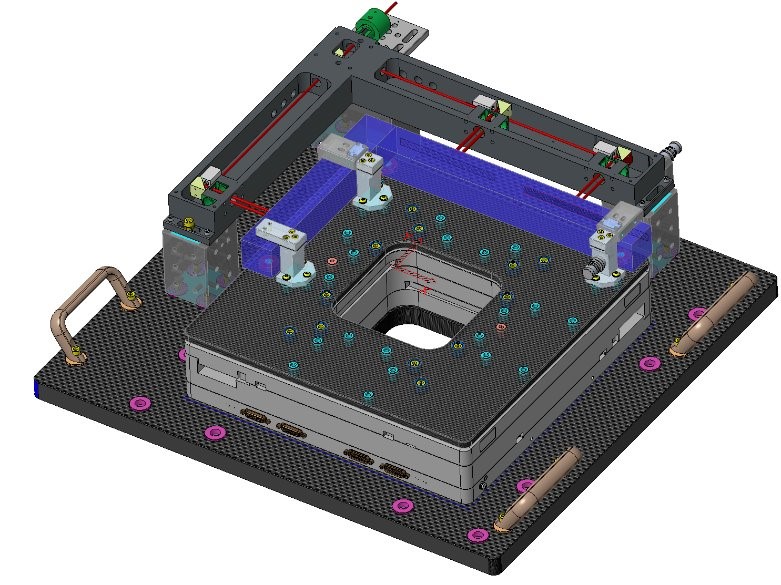}
   \end{tabular}
   \end{center}
   \caption[photo] 
   { \label{fig:photo} 
A two-axis system according to \ref{fig:block}b after finished assembly during adjustment against a L-shaped mirror (a) and model of XY stage system according to \ref{fig:block}a for optical metrology with three-axis system (b)}
\end{figure} 

For testing the receiver, a simple individual interferometer has been set up with a PZT element holding the mirror. Dedicated electronics control the linear scanning motion and digitise the output of the receiver under test. A designated software package evaluates and visualises the output in real-time, directly aiding the receiver alignment. 
For the static alignment, the testbed uses a precise L-shaped mirror to adjust the individual interferometers against it. For the dynamic alignment, a similar mirror is fixed on a precise XY translation stage (see example in Figure \ref{fig:photo} a)) and a testing electronic rig records data from all interferometers, environmental sensors and the stage itself during the predefined test patterns. Another dedicated software package controls the stage motion and data recording. The usual patterns include: linear motion in a single axis (X- and Y-), linear diagonal motion and linear motion with different velocity profiles. The patterns vary according to particular applications. Also, when specific mechatronics is part of the application (Figure \ref{fig:photo} b)), the dynamic testing could be carried out in situ.

\subsection{Testing and verification metrics}
\label{metrics}

For the intermittent testing and verification, we have defined a set of quantitative metrics that reflect the individual aspects related to the measurement performance of the interferometric systems. As mentioned above, we detect the imperfections and misalignments of the optical system by assessing its output. Specifically, we investigate the phase diagram at the output of the interferometric receivers. 
To quantify the individual metrics, we use well-established techniques for compensating the linearity errors\cite{cip2000scaleLinearization}, based on ellipse-fitting, in a reverse manner, i.e. to characterize the residual cyclic error from the phase diagram.

\begin{figure} [ht]
   \begin{center}
   \begin{tabular}{c} 
   \includegraphics[width=10cm]{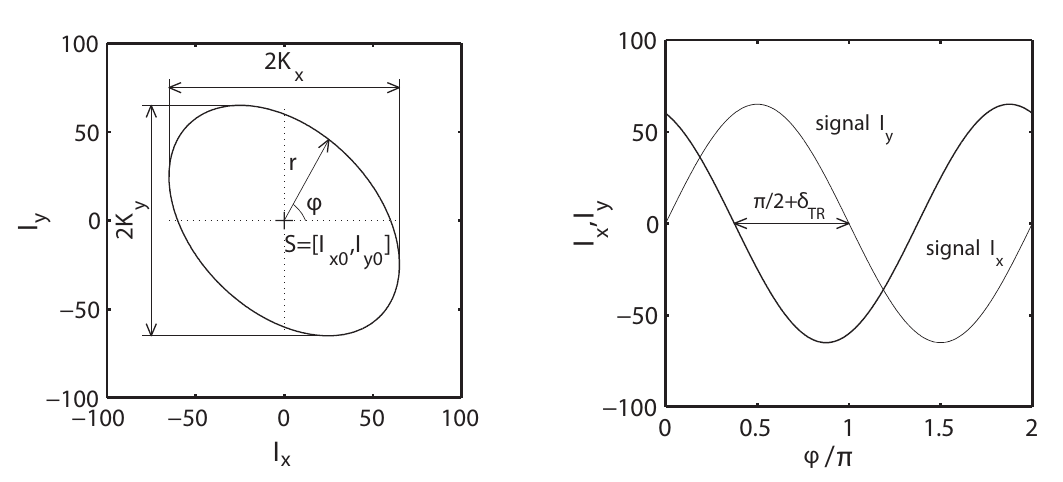}
   \end{tabular}
   \end{center}
   \caption[example] 
   { \label{fig:nonlin} The Lissajous phase diagram parameters visualized (adopted from \cite{cip2000scaleLinearization})}
\end{figure} 

The parameters of the phase diagram reflect the quality of the interferometer output in terms of deviations from an ideal signal. Ideally, when the quadrature signal from the detection unit is plotted on a Lissajous phase diagram as a function of movement, it traces a circle centred at the origin. From a signal perspective, both components of the quadrature signal, $I_x$ (cosine) and $I_y$ (sine), should have identical amplitudes ($K_x = K_y$), zero offsets ($I_{x0} = I_{y0} = 0$), and a phase difference of exactly $\pi/2$.

In practice, deviations from this ideal behaviour occur, as indicated in Figure \ref{fig:nonlin}, meaning $K_x \neq K_y$, $I_{x0} \neq I_{y0} \neq 0$, and the phase difference between $I_x$ and $I_y$ is $\pi/2 + \beta$, where $\beta$ is the residual phase error between the two components of the quadrature. All these deviations contribute to increased periodic non-linearity in the interferometric measurement. For testing purposes, we investigate for three primary parameters: \textit{Egginess} (denoted $Z_m$), calculated as
\begin{equation}
    Z_m = \frac{(K_x / K_y)}{K_z}, \quad \text{where} \quad K_z = \frac{(K_x + K_y)}{2};
\end{equation}
\textit{Miscentering} ($Z_0$), which is the square rood of the sum of squares of the individual offsets $I_{x0}$ and $I_{y0}$ normalised to the average amplitude $K_z$, i.e.,
\begin{equation}
    Z_0 = \frac{\sqrt{(I_{x0} ^ 2 + I_{y0} ^ 2)}}{K_z};
\end{equation}
and \textit{Phase Skew} ($\beta$), the detected phase deviation between X and Y, given by
\begin{equation}
    \beta = \phi(I_X) - \phi(I_y) - \frac{\pi}{2}.
\end{equation} for the momentary phase $\phi$.
Egginess primarily stems from component tolerances/non-linearities in the preamplifier and subsequent imperfect gain adjustments in the receiver unit. Miscentering is mainly due to component tolerances in the preamplifier. Phase skew is primarily caused by imperfections in and the adjustment of the detection unit's quarter-wave plate (QWP). The \textit{Overall Non-linearity} ($NL$) is the calculated maximum deviation from the mean value (peak-to-peak), which is converted to nanometers.

Other parameters related to the behaviour of the quadrature signal amplitude as a function of the test mirror's movement range: \textit{Potency} ($H$), the gain coefficient representing the average amplitude of the quadrature signals (X, Y) in Volts divided by the laser beam power ($P_{in}$) at the input (the required value is customer-specified); \textit{Ripple} ($R$ or $\rho$), expressed as twice the standard deviation of the amplitude fluctuations during the test ($2\sigma(\rho)$) normalized to amplitude. Ripple is influenced by errors in the mirror stage guidance, imperfect adjustment of the tested corner cube, and the phase diagram offset (non-zero offset). Finally, the \textit{Contrast Fade} ($R_s$) is calculated from a linear fit of the Ripple, representing the amplitude variation during the test as a function of the stage position along the relevant interferometer axis, expressed as the ratio of the difference between the maximum and minimum measured values to the average amplitude over the entire test. Besides guidance errors, the contrast fade is mainly caused by geometric adjustment errors of either the interferometer itself within the corner cube or the corner cube at the testing station.

\subsection{Sample Results}
\label{results}

\begin{figure} [ht]
   \begin{center}
   \begin{tabular}{c} 
   \includegraphics[width=\textwidth]{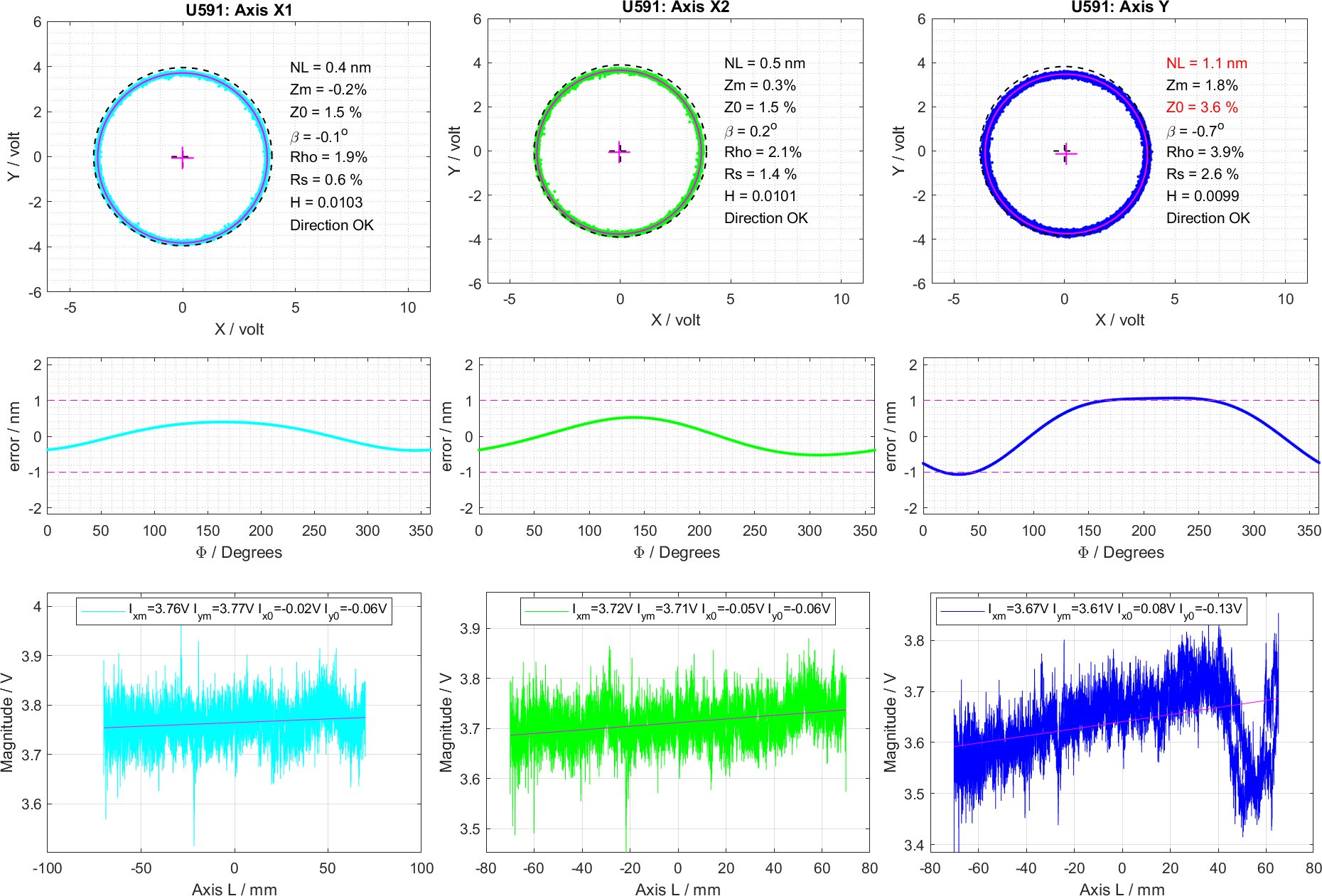}
   \end{tabular}
   \end{center}
   \caption[example] 
   { \label{fig:sampleout} 
A sample verification result visualised -- the three columns correspond to three interferometric axes, where each column displays the Lissajous phase plot with the nominal course and metrics indicated (top), the cyclic error plot (middle) and the interferometric contrast plot over the displacement (bottom).}
\end{figure} 


To illustrate the verification methodology, we present representative data obtained from a three-axis vingel system incorporating two X-axis and one Y-axis interferometers. Dynamic diagonal scans over a $150\,$mm travel range in both axes were performed.

The verification result, visualised in Figure \ref{fig:sampleout}, indicates a total cyclic error less than $0,5\,$nm for the X interferometers. The phase diagram of the Y interferometer shows a noticeable offset from the centre, resulting in increased periodic non-linearity and indicating the need for re-adjustmen

\section{FINAL WORDS}
\label{final}




We have presented the development, assembly, and systematic verification of a compact monolithic interferometric system — \textit{the vingel} — designed for high-precision coordinate positioning in demanding industrial and scientific applications. The monolithic architecture, incorporating two-axis and three-axis configurations, enables sub-nanometer resolution while maintaining long-term stability, compact form factor, and compatibility with ultra-high vacuum environments.

A key advantage of the proposed design is the complete pre-adjustment of optical components during assembly, reducing the burden of on-site alignment and ensuring repeatable performance. The integration of multiple interferometric axes into a single rigid structure inherently minimizes geometric errors and thermally induced drifts, which are critical in applications such as nanometrology, e-beam lithography, and semiconductor manufacturing.

The verification protocol, described in Section~\ref{metrics}, introduces a comprehensive set of diagnostic metrics derived from quadrature signal analysis, including phase diagram parameters, cyclic errors, and dynamic ripple behavior. These metrics allow for quantitative assessment of system performance at each stage of the assembly and provide actionable feedback for continuous refinement of the production process. The presented verification results (Section~\ref{results}) illustrate the diagnostic capability to identify residual alignment errors in the Y-axis channel. The employed methodology offers a robust framework for both quality assurance and design optimization.

Overall, the vingel system contributes a scalable and manufacturable solution to the growing demands of high-accuracy displacement measurement in precision engineering fields. Future work will focus on extending the approach to additional degrees of freedom, integration into complex mechatronic platforms, and further automation of the assembly and verification processes.

\section*{DISCLOSURE}

The authors declare no conflicts of interest. 
During the preparation of this manuscript, the authors used AI-based tools including Grammarly (grammarly.com), Perplexity (perplexity.ai), Gemini (gemini.google.com), and ChatGPT (openai.com) for particular linguistic and structural refinement. All content was subsequently reviewed, edited, and verified by the authors, who assume full responsibility for the final text.


\acknowledgments 

The authors acknowledge financial support from the Technology Agency of the Czech Republic (project TN02000020) and institutional support from the Czech Academy of Sciences (RVO:68081731).

\bibliography{report} 
\bibliographystyle{spiebib} 

\end{document}